\documentclass[amssymb,pra,twocolumn,notitlepage,superscriptaddress]{revtex4-2}
\pdfoutput=1
\usepackage{times}
\usepackage{amsmath}
\usepackage{ulem}
\usepackage{graphicx}
\usepackage{dsfont}
\usepackage{color}

\usepackage{braket}
\usepackage[colorlinks=true, letterpaper=ture, pdfstartview=FitV, linkcolor=blue, citecolor=blue, urlcolor=blue]{hyperref}
\usepackage{amssymb}
\usepackage{enumitem}
\usepackage{mathtools}
\usepackage{mathrsfs}
\usepackage[most]{tcolorbox}
\usepackage[margin=1in]{geometry}
\begin{document}
	\title{Adaptive random compiler for Hamiltonian simulation}
	\author{Yun-Zhuo Fan}
	\affiliation {Key Laboratory of Atomic and Subatomic Structure and Quantum Control (Ministry of Education), Guangdong Basic Research Center of Excellence for Structure and Fundamental Interactions of Matter, School of Physics, South China Normal University, Guangzhou 510006, China}
	\author{Yu-Xia Wu}
	\affiliation {Key Laboratory of Atomic and Subatomic Structure and Quantum Control (Ministry of Education), Guangdong Basic Research Center of Excellence for Structure and Fundamental Interactions of Matter, School of Physics, South China Normal University, Guangzhou 510006, China}
	\author{Dan-Bo Zhang}
	\email{dbzhang@m.scnu.edu.cn}
	\affiliation {Key Laboratory of Atomic and Subatomic Structure and Quantum Control (Ministry of Education), Guangdong Basic Research Center of Excellence for Structure and Fundamental Interactions of Matter, School of Physics, South China Normal University, Guangzhou 510006, China}
	\affiliation {Guangdong Provincial Key Laboratory of Quantum Engineering and Quantum Materials, Guangdong-Hong Kong Joint Laboratory of Quantum Matter, Frontier Research Institute for Physics, South China Normal University, Guangzhou 510006, China}
	\date{\today}
	
	\begin{abstract}
		
		Randomized compilation protocols have recently attracted attention as alternatives to traditional deterministic Trotter-Suzuki methods, potentially reducing circuit depth and resource overhead. These protocols determine gate application probabilities based on the strengths of Hamiltonian terms, as measured by the Schatten-$\infty$ norm. However, relying solely on the Schatten-$\infty$ norm to define sampling distributions may not be optimal, especially for continuous-variable and hybrid-variable systems involving unbounded operators, where quantifying Hamiltonian strengths is challenging. In this work, we propose an adaptive randomized compilation algorithm that dynamically updates sampling weights via low-order moment measurements of Hamiltonian terms, assigning higher probabilities to terms with greater uncertainty. This approach improves accuracy without significantly increasing gate counts and extends randomized compilation to continuous-variable and hybrid-variable systems by addressing the difficulties in characterizing the strengths of unbounded Hamiltonian terms. Numerical simulations demonstrate the effectiveness of our method.
		
	\end{abstract}
	
	\maketitle
	
	\section{introduction}
	Quantum computing offers a new way to solve problems that are difficult or even impossible for classical computers, especially in simulating quantum systems~\cite{Feynman1982simulating,Deutsch1985quantum,Lloyd1996universal,Andrew1998quantum}. One of the most important tasks in this area is Hamiltonian simulation~\cite{Childs2012hamiltonian,Low2017optimal,Low2019hamiltonian,Csahinouglu2021hamiltonian}, which involves using a quantum computer to reproduce the time evolution of a quantum system governed by a given Hamiltonian. This task is not only central to understanding physics itself~\cite{Somma2003quantum,Zohar2016quantum,Hartmann2016quantum,Bauer2023quantum} but also supports many practical applications, including quantum chemistry~\cite{Aspuru2005simulated,Babbush2014adiabatic,Hempel2018quantum,Arguello2019analogue,Li2019variational}, condensed matter physics~\cite{Zhang2011experomental,Hague2014cold,Tarruell2018quantum,Hofstetter2018quantum}, and quantum algorithms such as phase estimation~\cite{D1998general,Dorner2009optimal,Paesani2017experimental,Liu2021distributed,Smith2024adaptive} or digitized adiabatic quantum computing~\cite{Barends2016digitized,Cui2020circuit,Hegade2021shortcuts}.
	
	A Hamiltonian generally consists of multiple terms $H=\sum_{j}H_j$, where each term may describe interactions between particles, couplings between the system and external fields, or other relevant physical effects. Simulating such a Hamiltonian requires approximating the evolution operator $e^{-iHt}$ with high accuracy while minimizing the number of quantum gates, since current quantum hardware is limited by coherence time and circuit depth~\cite{Jeremy2007optical,Preskill2018quantum,Gyongyosi2019survey,Bruzewicz2019trapped,Kjaergaard2020superconducting}.
	
	One of the earliest and most widely used strategies for Hamiltonian simulation is the Trotter-Suzuki decomposition~\cite{Trotter1959product,Suzuki1976generalized,De1983applications,Suzuki1993improved,Zhao2023making,Zhao2025entanglement}. It works by first breaking the total time evolution $e^{-iHt}$ into $N$ steps of short-time evolution, with each step covering a small time interval $\Delta t=t/N$. The short-time evolution $e^{-iH\Delta t}$ is then approximated by a product of exponentials of individual terms in the Hamiltonian, $\prod_{j}e^{-iH_j\Delta t}$. This method is therefore also known as the product formula approach~\cite{Chorin1978product,De1987product,Huyghebaert1990product,Tran2020destructive,Chen2022efficient}. The rigorous bounds on the circuit size depend on the number of terms $L$ in the system Hamiltonian and the size of the largest term in the Hamiltonian. Consequently, the Trotter-Suzuki method is only practical for sparse Hamiltonians. While the Trotter-Suzuki decomposition is a deterministic compilation method, recent studies have shown that randomized compiling~\cite{Campbell2019random,Ouyang2020compilation,Chen2021concentration,Nakajj2024high,David2025tighter} has the potential to offer lower overheads.
	
	Random compiler~\cite{Campbell2019random} approximates evolution by randomly applying operations corresponding to Hamiltonian terms. These methods do not follow a fixed order, but randomly select terms based on the probability distribution of Hamiltonian term strengths. This technology reduces circuit depth and simplifies implementation, while maintaining controllable expected simulation errors. The key advantage of this strategy is that the circuit complexity depends on the absolute sum of the Hamiltonian term strengths (the Schatten-$\infty$ norm),  $\lambda=\sum_{j}\left|\left|H_j\right|\right|_\infty$,  rather than the number of terms $L$ or the size of the largest term in the Hamiltonian. In practice, this has led to meaningful improvements in both the accuracy and efficiency of simulations, particularly for electronic structure Hamiltonians~\cite{Whitfield2011simulation,Kivlichan2018quantum,Tranter2019ordering,Shee2022qubit} relevant to quantum chemistry. However, this algorithm relies heavily on using the Schatten-$\infty$ norm to determine the Hamiltonian strength, thereby obtaining a fixed sampling distribution, which may not be optimal for some situations. For example, it does not take into account how the quantum state evolves over time, nor does it adapt to terms that might temporarily dominate the dynamical behavior of the system. In particular, when dealing with continuous-variable systems~\cite{Duan2000inseparability,Simon2000peres,Adesso2007entanglement}, such as those found in photonic~\cite{Duan2004scalable,Kok2007linear,Slussarenko2019photonic}, trapped-ion~\cite{Gulde2003implementation,Kielpinski2002architecture,Haffner2008quantum} and superconducting quantum computing~\cite{Gambetta2017building,Huang2020superconducting,Wu2021strong}, the Hamiltonian may involve unbounded operators, making it difficult to use norm-based sampling rules.
	
	In this work, we propose an adaptive randomized simulation algorithm that improves upon fixed strategies by introducing a adaptive mechanism. At each step of the simulation, we measure low-order properties of each Hamiltonian term. These measurements guide the algorithm to update the sampling probabilities dynamically, giving more attention to the terms that are currently more important. This method retains the advantages of randomness and enables the algorithm to dynamically adapt to changes in the quantum system. Another contribution of our work is to extend this idea beyond discrete qubit systems to continuous-variable systems. Our method estimates moments of Hamiltonians instead of relying on fixed norms, providing a new way to define sampling weights, even in infinite-dimensional settings.
	
	This paper is organized as follows. In Sec.~\ref{section 2}, we present the framework of the adaptive random compiler, detailing its measurement strategy. In Sec.~\ref{section 3}, we demonstrate the effectiveness of the proposed method through numerical simulation across three system types: discrete-variable, continuous-variable, and hybrid-variable quantum systems. Finally,  Sec.~\ref{section 4} concludes with a discussion of our key findings and their implications.
	
	\section{adaptive random compiler for hamiltonian simulation}\label{section 2}
	
	In this section, we first introduce the adaptive random compiler framework. Then we show the measurement strategy for implementing this algorithm.
	
	\subsection{Adaptive random compiler}
	For the Hamiltonian simulation problem, the first step typically involves decomposing the Hamiltonian into multiple terms, which may be local in some cases. Without loss of generality, we consider a Hamiltonian that can be decomposed into the following form:
	\begin{eqnarray}\label{Eq.(r1)}
		H=\sum_{j}^{L}H_j=\sum_{j}^{L}h_j\frac{H_j}{h_j}=\sum_{j}^{L}h_jH'_j.
	\end{eqnarray}
	We can always choose $H'_j=H_j/h_j$ so that the weighting $h_j$ is positive real number.
	
	In the original protocol of random compiler~\cite{Campbell2019random}, the weighting $h_j$ is obtained by normalizing $H_j$, such that the largest singular value of $H'_j$ is 1. Based on this weighting, a probability distribution $p_j=h_j/(\sum_{k}h_k)$ is computed, which is then independently sampled to generate the entire sequence of quantum gates to be applied. Due to a bias introduced into the probability distribution, the evolution statistically approaches the target unitary after many repetitions.
	
	\begin{figure}[htbp]
		\begin{tcolorbox}[colframe=black, colback=white, sharp corners, fonttitle=\bfseries, title=Adaptive random compiler]
			
			\begin{tabular}{@{}l p{0.95\linewidth}@{}}
				&\textbf{Input:} A list of Hamiltonian terms \( H = \sum_j H_j \), total evolution time \( t \), number of evolution steps \( N \), the initial state \( \ket{\psi_{0}} \) and a classical oracle function SAMPLE() that returns an index \( j \) from a given probability distribution \( \left\{p_j\right\} \). \\
				&\textbf{Output:} The final state \( \ket{\psi_{N}} \) after \( N \) steps of evolution.
			\end{tabular}
			
			\begin{enumerate}[label=\arabic*.]
				\item Repeat steps 2-6 for \( k=0, 1, ..., N-1 \).
				\item For the current input state \( \ket{\psi_{k}} \), measure the expectation values of low-order moments for all terms in the Hamiltonian, \( \braket{H_j}\), \( \braket{H^2_j}\), \( \braket{H^3_j}\), and \( \braket{H^4_j}\).
				\item Use the measurement results from the previous step to obtain the optimal probability distribution: \(p_j=\frac{(6\braket{H_j^2}^2-8\braket{H_j}\braket{H_j^3}+2\braket{H_j^4})^{1/4}}{\sum_{k}(6\braket{H_k^2}^2-8\braket{H_k}\braket{H_k^3}+2\braket{H_k^4})^{1/4}}\).
				\item Sample an index \( j \gets\) SAMPLE() according to the distribution \( \left\{p_j\right\} \).
				\item Calculate the time slice: \(\tau_j=\frac{t}{Np_j}\).
				\item Update the state by applying the unitary gate: \( \ket{\psi_{k+1}}=e^{-i\tau_j H_j}\ket{\psi_{k}} \).
				\item \textbf{Return} \( \ket{\psi_{N}} \).
			\end{enumerate}
			
		\end{tcolorbox}
		\caption{Pseudocode for adaptive random compiler protocol.\label{Fig.1}}
	\end{figure}
	
	However, in this work, $p_j$ is not fixed, but rather treated as a variational parameter optimized dynamically, with the goal of adapting the sampling probability distribution at each step to achieve improved simulation performance. Our full algorithm is given as pseudocode in Fig.~\ref{Fig.1}. At each step, we perform a set of expectation value measurements on the state from the previous step to obtain the cost function $\epsilon\left(\left\{p_j\right\}\right)$, which quantifies the error between the evolution under randomized compiling and the exact evolution. Due to the convexity and separability of the cost function, the optimal probability distribution $\left\{p_j\right\}$ can be analytically derived using the method of Lagrange multipliers~\cite{Nocedal1999numerical,Boyd2004convex}. This distribution is subsequently sampled to decide which unitary gate $e^{-i\tau_j H_j}$ to apply next, where $\tau_j=\frac{t}{Np_j}$.
	
	\subsection{Probability distribution}
	
	In the following, we will show how to obtain the optimal probability distribution $\left\{p_j\right\}$. Since the unitary applied at each step is sampled from a probability distribution, the evolution of each step is mathematically represented by a quantum channel as follows
	\begin{eqnarray}
		\mathcal{E}(\rho)&=&\sum_{j}\frac{h_j}{\sum_{k}h_k}e^{-iH_j\tau_j}{\rho}e^{iH_j\tau_j}\nonumber\\
		&=&\sum_{j}p_je^{-iH_j\tau_j}{\rho}e^{iH_j\tau_j}\nonumber\\
		&=&\sum_{j}p_je^{\tau_j\mathcal{L}_j}(\rho).
		\label{Eq.2}
	\end{eqnarray}
	where $\mathcal{L}_j$ is the Liouvillian superoperator, defined by $\mathcal{L}_j( \rho)=-i[H_j,\rho]$. Similarly, we can use the Liouvillian representation to express the exact time evolution. For each step, we can write it in the following form:
	\begin{eqnarray}
		\mathcal{U}_N(\rho)=e^{-iHt/N}{\rho}e^{iHt/N}=e^{t\mathcal{L}/N}(\rho).
		\label{Eq.3}
	\end{eqnarray}
	where $t$ is the total evolution time, and $N$ denotes the number of steps. Moreover, we have that $\mathcal{L}=\sum_{j}\mathcal{L}_j$. By expanding Eq.~\eqref{Eq.2} and Eq.~\eqref{Eq.3}, we can obtain
	\begin{eqnarray}
		\mathcal{E}&=&\mathds{1}+\left(\sum_{j}p_j\tau_j\mathcal{L}_j\right)+\sum_{j}p_j\sum_{n=2}^{\infty}\frac{\tau_j^n\mathcal{L}_j^{n}}{n!}.
	\end{eqnarray}
	\begin{eqnarray}
		\mathcal{U}_N=\mathds{1}+\frac{t}{N}\mathcal{L}+\sum_{n=2}^{\infty}{\frac{t^n\mathcal{L}^n}{n!N^n}}.
	\end{eqnarray}
	We see the first two terms of $\mathcal{E}$ and $\mathcal{U}_N$ match whenever $\tau_j=\frac{t}{Np_j}$. Using this value for $\tau_j$ and taking the large $N$ limit, we quantify the difference between two quantum channels by the Hilbert--Schmidt norm $||A||=\sqrt{\mathrm{Tr} (A^\dagger A)}$. We then have
	\begin{eqnarray}
		\left|\left|\mathcal{U}_N(\rho)-\mathcal{E}(\rho)\right|\right|\approx\left|\left|\frac{t^2\mathcal{L}^2(\rho)}{2N^2}-\sum_{j}\frac{ t^2\mathcal{L}_j^{2}(\rho)}{2N^2p_j}\right|\right|.
	\end{eqnarray}
	We further simplify the expression using the following notation:
	\begin{eqnarray}
		D(\rho)=\mathcal{L}^2(\rho)=-\left[H,\left[H,\rho\right]\right].
	\end{eqnarray}
	\begin{eqnarray}
		D_{ij}(\rho)=\mathcal{L}_i\mathcal{L}_j(\rho)=-\left[H_i,\left[H_j,\rho\right]\right].
	\end{eqnarray}
	We thus obtain
	\begin{eqnarray}
		&&\left|\left|\frac{t^2\mathcal{L}^2(\rho)}{2N^2}-\sum_{j}\frac{ t^2\mathcal{L}_j^{2}(\rho)}{2N^2p_j}\right|\right|\nonumber\\
		&&=\frac{t^2}{2N^2}\left|\left|D(\rho)-\sum_{j}\frac{ D_{jj}(\rho)}{p_j}\right|\right|\nonumber\\
		&&\leq\frac{t^2}{2N^2}\left(\left|\left|D(\rho)\right|\right|+\sum_{j}\frac{\left|\left|D_{jj}(\rho)\right|\right|}{p_j}\right).
	\end{eqnarray}
	The above expression allows us to determine which set of probabilities ${p_j}$ results in a smaller difference between the two quantum channels. Moreover, the entire expression above is not necessary to serve as the cost function, since the term $\left|\left|D(\rho)\right|\right|$ does not contain any parameters and is therefore a constant. Thus, for a given density matrix $\rho$, the cost function can be defined as follows
	\begin{eqnarray}
		\epsilon(\left\{p_j\right\})=\sum_{j}\frac{\left|\left|D_{jj}(\rho)\right|\right|}{p_j}.
	\end{eqnarray}
	Each term of this cost function is an inverse proportional function of $p_j$, and since all probabilities are non-negative, the problem exhibits strong convexity and is analytically tractable. Therefore, besides directly optimizing it using classical optimizers, we can also solve such problems using the method of Lagrange multipliers~\cite{Nocedal1999numerical,Boyd2004convex}. To minimize the cost function under the constraint condition $\sum_{j}p_j=1$, we introduce the Lagrange multiplier $\mu$ and define the Lagrangian function in the following form:
	\begin{eqnarray}
		\mathscr{L}=\sum_{j}\frac{\left|\left|D_{jj}(\rho)\right|\right|}{p_j}+\mu\left(\sum_{j}p_j-1\right).
	\end{eqnarray}
	Then, we take the derivative of the Lagrangian function to find its extremum point.
	\begin{eqnarray}
		\frac{\partial\mathscr{L}}{\partial p_j}=-\frac{\|D_{jj}(\rho)\|}{p_j^2}+\mu=0.
	\end{eqnarray}
	Thus, we obtain the optimal probability distribution.
	\begin{eqnarray}
		p_j=\frac{\sqrt{\|D_{jj}(\rho)\|}}{\sum_k \sqrt{\|D_{kk}(\rho)\|}}.
		\label{Eq.13}
	\end{eqnarray}
	From the above equation, we can see that the probability $p_j$ is proportional to $\sqrt{\|D_{jj}(\rho)\|}$, indicating that Hamiltonian terms with greater uncertainty are assigned higher probabilities.
	
	So far, we have established that the optimal probability distribution $\left\{p_j\right\}$ can be obtained by measuring $\left\{\left|\left|D_{jj}(\rho)\right|\right|\right\}$. We will introduce the measurement strategy in the next part.
	
	\subsection{Measurement strategy}
	Let us first consider general quantum states which are described by density matrix. By using the Taylor expansion of the time evolution operator, we can obtain
	\begin{eqnarray}
		&&\left|\left|D_{jj}(\rho)\right|\right|=\left|\left| \left[H_j,\left[H_j,\rho\right]\right] \right|\right|\nonumber\\
		&&=\lim_{\Delta t\rightarrow 0}\frac{\left|\left|e^{iH_j\Delta t}\rho e^{-iH_j\Delta t}+e^{-iH_j\Delta t}\rho e^{iH_j\Delta t}-2\rho \right|\right|}{(\Delta t)^2}.\nonumber\\
	\end{eqnarray}
	We can use the following notations to simplify the expression.
	\begin{eqnarray}
		\rho_1=e^{iH_j\Delta t}\rho e^{-iH_j\Delta t}.
	\end{eqnarray}
	\begin{eqnarray}
		\rho_2=e^{-iH_j\Delta t}\rho e^{iH_j\Delta t}.
	\end{eqnarray}
	According to the definition of the Hilbert-Schmidt norm and the Hermitian property of the density matrix, we can obtain
	\begin{eqnarray}
		&&\left|\left|\rho_1+\rho_2-2\rho \right|\right|
		=\sqrt{\mathrm{Tr}\left[\left(\rho_1+\rho_2-2\rho\right)^2\right]}\nonumber\\
		&&=\sqrt{\mathrm{Tr}\left(\rho_1^2+\rho_2^2+4\rho^2+2\rho_1\rho_2-4\rho_1\rho-4\rho_2\rho\right)}.\nonumber\\
	\end{eqnarray}
	where $\mathrm{Tr}\left(\rho_1^2\right)$, $\mathrm{Tr}\left(\rho_2^2\right)$ and $\mathrm{Tr}\left(\rho^2\right)$ are the purities of the three corresponding density matrices, respectively. $\mathrm{Tr}\rho_1\rho_2$, $\mathrm{Tr}\rho_1\rho$ and $\mathrm{Tr}\rho_2\rho$ represent the Hilbert-Schmidt inner products between pairs of density matrices, which quantify the similarity between them. After determining these values via the SWAP test~\cite{Buhrman2001quantum} or randomized measurements~\cite{Elben2019statistical,Elben2020cross}, we can obtain $\left|\left|D_{jj}(\rho)\right|\right|$.
	
	For pure states $\rho=\ket{\psi}\bra{\psi}$, measurements of $\left|\left|D_{jj}(\rho)\right|\right|$ can be greatly simplified. According to the definition of the Hilbert-Schmidt norm together with the Hermitian properties of the density matrix and the Hamiltonian, we can obtain
	\begin{eqnarray}
		\left|\left|D_{jj}(\rho)\right|\right|&=&\left|\left| \left[H_j,\left[H_j,\rho\right]\right] \right|\right|\nonumber\\
		&=&\sqrt{\mathrm{Tr}\left[\left(H_j^2\rho-2H_j\rho H_j+\rho H_j^2\right)^2\right]}.\nonumber\\
		&=&\sqrt{6\braket{H_j^2}^2-8\braket{H_j}\braket{H_j^3}+2\braket{H_j^4}}.\nonumber\\
		\label{Eq.18}
	\end{eqnarray}
	Therefore, by measuring the low-order moments of $H_j$, we can determine $\left|\left|D_{jj}(\rho)\right|\right|$. By substituting Eq.~\eqref{Eq.18} into Eq.~\eqref{Eq.13}, we can obtain the optimal probability distribution,
	\begin{eqnarray}
		p_j=\frac{(6\braket{H_j^2}^2-8\braket{H_j}\braket{H_j^3}+2\braket{H_j^4})^{\frac{1}{4}}}{\sum_{k}(6\braket{H_k^2}^2-8\braket{H_k}\braket{H_k^3}+2\braket{H_k^4})^{\frac{1}{4}}}.
	\end{eqnarray}
	
	It should be noted that although each projective measurement collapses the quantum state and requires repreparation of the state, in both mixed state~\cite{Buhrman2001quantum,Elben2019statistical,Elben2020cross} and pure state~\cite{Aulicino2022state,Huang2020predicting} cases only a polynomial number of measurements is needed. Moreover, since we applied the method of Lagrange multipliers~\cite{Nocedal1999numerical,Boyd2004convex} to the cost function, no additional cost will be introduced in the classical post-processing stage.
	
	\subsection{Analysis of complexity}
	\begin{table*}[htbp]
		\caption{Comparison of resources used by the first-order Trotter, random compiler (RC), and adaptive random compiler (ARC) methods. Here, $L$ is the number of Hamiltonian terms, $\Lambda$ is the size of the largest term in the Hamiltonian, $\lambda$ is the absolute sum of Hamiltonian strengths, $t$ is the total evolution time, $\epsilon$ denotes the upper bound of the Hamiltonian simulation error, and $\mathcal{L}_j$ is the Liouvillian superoperator defined as $\mathcal{L}_j(\rho) = -i[H_j,\rho]$, with $\mathcal{L}=\sum_j\mathcal{L}_j$. The symbols $\|\cdot\|$ and $\|\cdot\|_{\infty}$ denote the Hilbert--Schmidt norm and the Schatten-$\infty$ norm, respectively. $\rho_i$ represents the exact evolved density matrix at the $i$th step, and $\overline{(\cdot)}$ indicates the average over all evolution steps. $\varepsilon$ denotes the upper bound on the statistical error of the measurement outcomes. $k$ and $w$ denote the maximum number of qubits nontrivially acted upon by any Pauli string obtained from expanding the measured observables. $k$ corresponds to strings from observables measured in real-time dynamics, and $w$ corresponds to strings from the adaptive measurement. $N$ is the total number of qubits, while $O(N^{S})$ and $O(N^{4R})$ denote the scaling of the number of Pauli strings obtained from expanding the measured observables in real-time dynamics and in adaptive measurements, respectively.\label{table_1}}
		\begin{ruledtabular}
			\begin{tabular}{p{2.1cm} c c c c}
				Protocol & \shortstack{Circuit depth \\ (state-independent)} 
				& \shortstack{Circuit depth \\ (state-dependent)} 
				& \shortstack{Shots per evolution step \\ (state preparation)} 
				& \shortstack{Shots per evolution step \\ (real-time dynamics)} \\
				\hline
				1st Trotter~\cite{Trotter1959product} 
				& $O[\frac{L^3 (\Lambda t)^2}{\epsilon}]$
				& $O[\frac{Lt^2\overline{||\sum_{j<k}[\mathcal{L}_j,\mathcal{L}_k](\rho_i)||}}{\epsilon}]$
				& 0 
				&$\Omega(\frac{3^kS\log N}{\varepsilon^2})$ \\[3pt]
				
				RC~\cite{Campbell2019random}
				& $O[\frac{(\lambda t)^2}{\epsilon}]$
				& $O[\frac{t^2\overline{(||\mathcal{L}^2(\rho_i)||+\lambda\sum_{j}\frac{||\mathcal{L}_j^2(\rho_i)||}{||H_j||_{\infty}})}}{\epsilon}]$
				& 0 
				& $\Omega(\frac{3^kS\log N}{\varepsilon^2})$ \\[3pt]
				
				ARC (this work)
				& N/A 
				& $O[\frac{t^2\overline{(||\mathcal{L}^2(\rho_i)||+(\sum_{j}\sqrt{||\mathcal{L}_j^2(\rho_i)||})^2)}}{\epsilon}]$
				&$\Omega(\frac{3^w(4R)\log N}{\varepsilon^2})$
				& $\Omega(\frac{3^{\max\{w,k\}}\max\{S,4R\}\log N}{\varepsilon^2})$ \\[3pt]
			\end{tabular}
		\end{ruledtabular}
	\end{table*}
	
	We give a comparison of complexity for the first-order Trotter formula~\cite{Trotter1959product} , original random compiler (RC)~\cite{Campbell2019random}, and our adaptive random compiler (ARC), which is shown in Table \Ref{table_1}. The analysis is as follows. In terms of state-independent circuit depth, the first-order Trotter formula exhibits a scaling of $O(L^3 (\Lambda t)^2 / \epsilon)$, while random compiling (RC) yields $O((\lambda t)^2 / \epsilon)$. Here, $t$ is the total evolution time, $\epsilon$ denotes the upper bound of the Hamiltonian simulation error. The adaptive random compiler (ARC), however, relies on measurements of the instantaneous state and corresponding adaptive selections during the algorithm, making it impossible to give a single, state-independent, universal expression for the circuit depth. Therefore, marking it as “N/A” in the table is appropriate. Intuitively, the first-order Trotter expression explicitly contains higher-order factors of the number of Hamiltonian terms $L$ and the size of the largest term in the Hamiltonian $\Lambda$, whereas the RC depth depends on the absolute sum of operator norms of the Hamiltonian terms $\lambda$, which can be significantly smaller in many realistic systems with long-range interactions. Consequently, under state-independent analysis, RC often outperforms the first-order Trotter approach.
	
	Comparisons based on state-dependent circuit depth provide a more faithful reflection of each algorithm behavior. The state-dependent upper bounds of the three methods take the following forms: the first-order Trotter has the scaling form $O[\frac{Lt^2\overline{||\sum_{j<k}[\mathcal{L}_j,\mathcal{L}_k](\rho_i)||^2}}{\epsilon}]$, RC has the scaling form $O[\frac{t^2\overline{(||\mathcal{L}^2(\rho_i)||+\lambda\sum_{j}\frac{||\mathcal{L}_j^2(\rho_i)||}{||H_j||_{\infty}})}}{\epsilon}]$, and ARC has the scaling form $O[\frac{t^2\overline{(||\mathcal{L}^2(\rho_i)||+(\sum_{j}\sqrt{||\mathcal{L}_j^2(\rho_i)||})^2)}}{\epsilon}]$. Here, $\mathcal{L}_j$ is the Liouvillian superoperator defined as $\mathcal{L}_j(\rho) = -i[H_j,\rho]$, with $\mathcal{L}=\sum_j\mathcal{L}_j$. The symbols $\|\cdot\|$ and $\|\cdot\|_{\infty}$ denote the Hilbert--Schmidt norm and the Schatten-$\infty$ norm, respectively. $\rho_i$ represents the exact evolved density matrix at the $i$th step, and $\overline{(\cdot)}$ indicates the average over all evolution steps. Note that the second term $\lambda\sum_{j}\frac{||\mathcal{L}_j^2(\rho_i)||}{||H_j||_{\infty}}$ in the numerator of RC and the corresponding term $(\sum_{j}\sqrt{||\mathcal{L}_j^2(\rho_i)||})^2$ in ARC can be compared via the Cauchy–Schwarz inequality $(\sum_{i}u_iv_i)^2\leq\left(\sum_{j}u_i^2\right)\left(\sum_{j}v_i^2\right)$ for $u_i,v_i\in\mathbb{R}$. 
	\begin{eqnarray}(\sum_{j}\sqrt{||\mathcal{L}_j^2(\rho_i)||})^2&&\leq(\sum_{j}||H_j||_{\infty})\left(\sum_{j}\frac{||\mathcal{L}_j^2(\rho_i)||}{||H_j||_{\infty}}\right)\nonumber\\
		&&=\lambda\sum_{j}\frac{||\mathcal{L}_j^2(\rho_i)||}{||H_j||_{\infty}}.
	\end{eqnarray}
	Thus, the corresponding term in ARC is no greater than that in RC. Since all other terms in both algorithms are identical, this inequality indicates that the upper bound of the state-dependent circuit depth in ARC is no larger than that of RC. In other words, ARC can achieve the same precision with a circuit depth no greater than that of RC, benefiting from its adaptive utilization of state information. The equality condition is both simple and stringent: it occurs only when, throughout the entire evolution, the probability distribution derived from operator norms in the original random compiling exactly coincides with the optimal adaptive probability distribution. This imposes strong constraints on both the Hamiltonian and the quantum state, and thus almost never occurs in practice. Therefore, ARC generally provides a strict improvement over RC, and equality of the upper bounds represents a practically unattainable special case. When the inequality is well satisfied, meaning that the adaptive probability distribution substantially deviates from the Schatten-$\infty$ norm distribution, ARC can reach the same target accuracy with a markedly smaller maximum circuit depth. Moreover, this advantage can become significant when the first term in the numerator is not dominant. In such cases, the benefit of adaptivity can become particularly evident.
	
	Directly comparing ARC with the first-order Trotter method is not straightforward from the table. Our approach is to first compare RC and the first-order Trotter in terms of their state-independent scaling, and then infer ARC advantage over the first-order Trotter indirectly through its improvement relative to RC. Specifically, in many realistic systems with long-range interactions, one typically finds $\lambda \ll \Lambda L$. Even when $\lambda \sim \Lambda L$, random compiling still clearly outperforms the first-order Trotter formula. Although RC exhibits a much better dependence on $L$, its scaling with respect to the evolution time $t$ is quadratic, whereas higher-order Trotter formulas (even the second-order one) exhibit a lower-order dependence on $t$. Hence, for a fixed Hamiltonian, RC performs better at short evolution times, but there always exists a critical time $t$ beyond which higher-order Trotter formulas become more efficient. Since ARC state-dependent circuit depth is no worse than that of RC, we infer that within the parameter regimes where RC outperforms the first-order Trotter method, ARC will inherit and amplify this advantage. However, when compared to higher-order product formulas, the advantage of ARC becomes less significant, as their better scaling with respect to $t$ allows them to outperform ARC beyond relatively short evolution times.
	
	In tasks targeting state preparation, the main drawback of ARC lies in the additional measurement shots required at each evolution step, which in turn increases the total computational resources. By contrast, both the first-order Trotter and RC require no extra measurements per step (denoted as 0 in the table), whereas ARC needs additional samples to estimate statistical quantities relevant to adaptive decisions and to adjust its strategy for the next step. The lower bound listed in the table is based on estimates using the classical shadow method~\cite{Huang2020predicting}. The motivation for adopting classical shadows is that when simultaneously estimating a large number of observables, the sample complexity of the shadow approach typically outperforms direct measurement, making it a reasonable lower bound on measurement cost. Specifically, one first expands the Hamiltonian into multiple Pauli strings, and then estimates these Pauli strings in parallel using classical shadows. Although the worst-case Hamiltonian can generate up to $4^N$ Pauli strings, realistic physical Hamiltonians are usually local, with a total Pauli string count scaling as $O(N^R)$, where $R\le4$. For instance, $R=1$ for the Heisenberg model, while $R=4$ corresponds to more complex electronic structure Hamiltonians, which typically involve only one- and two-body interactions. In these cases, the two-electron terms contribute $O(N^4)$ terms in the Hamiltonian~\cite{Wecker2014gate,Hastings2014improving}.  The scaling $\Omega(3^w(4R)\log N / \varepsilon^2)$ shown in the table reflects the lower bound on the measurement sample complexity under these structural constraints, where $w$ denotes the maximum number of qubits on which any of the previously expanded Pauli strings acts nontrivially, and $\varepsilon$ represents the upper bound on the statistical error of the measurement outcomes.
	
	%After mapping to qubits via encodings such as Jordan--Wigner~\cite{Jordan1928paulische}, parity~\cite{Seeley2012bravyi}, or Bravyi--Kitaev~\cite{Bravyi2002fermionic}, the resulting qubit Hamiltonian will similarly contain roughly $O(N^4)$ terms, where $N$ denotes the number of qubits, and each term is a Pauli string. Consequently, in the worst scaling scenario, the fourth-order moments may need to be expanded into a number of Pauli string combinations scaling as $O(N^{4R})$.
	
	Although ARC requires more measurement samples than the first-order Trotter and RC methods, resulting in an increase in overall resources, this does not imply that ARC is disadvantageous overall. Notably, the main bottleneck of NISQ devices usually lies in the limited circuit depth constrained by coherence time and gate noise, since deeper circuits lead to accumulated noise and significantly reduced fidelity. Therefore, under realistic hardware limitations, trading additional measurements for a reduction in circuit depth is a reasonable and often beneficial strategy. The depth reduction achieved through adaptive measurement in our algorithm directly improves the attainable precision, which may play a crucial role in demonstrating quantum advantage on NISQ devices.
	
	In real-time dynamics applications, ARC measurement overhead can be further mitigated. First, real-time dynamics inherently require measuring a set of observables at each time step (such as local spins, correlation functions, higher-order moments, or the Binder cumulant). If the Pauli strings required for these observables overlap substantially with those needed for ARC adaptive steps, the same measurement data can be reused, avoiding redundant sampling. Second, when the target observables in real-time dynamics are themselves complex, for example, when computing the Binder cumulant or high-order correlation functions to study phase transitions, the intrinsic measurement workload may already be comparable to or larger than that required by ARC adaptive estimation. In such cases, the relative additional cost of ARC becomes negligible. Consequently, in the study of real-time dynamics, critical behavior, or high-order statistical quantities, ARC can maintain a low circuit depth while keeping the measurement cost within a reasonable range. In Table \Ref{table_1}, we assume that the observables measured in real-time dynamics can be expanded into Pauli strings of order $O(N^S)$, with $k$ denotes the maximum number of qubits on which any of these Pauli strings acts nontrivially.
	
	In summary, RC provides a state-independent advantage over the first-order Trotter method in many realistic systems. ARC, as a state-dependent improvement upon RC, can be proven to yield a tighter upper bound on the state-dependent circuit depth. Under the realistic constraints of NISQ devices limited by circuit depth and noise, ARC strategy of trading measurement overhead for depth reduction proves practically valuable. In research scenarios involving extensive real-time measurements, the reuse of measurement data and the inherently high measurement complexity of the task often mitigate ARC adaptive overhead, rendering it competitive in overall resource efficiency.
	
	\section{simulation results}\label{section 3}
	In this section, we demonstrate the quantum algorithm using three model Hamiltonians involving discrete-variable, continuous-variable, and hybrid-variable systems. The numeral simulation is conducted using the open-source package \textit{QuTiP}~\cite{Lambert2024qutip,Johansson2013qutip,Johansson2012qutip}. Unless otherwise noted, all simulation results presented in Figs.~\ref{Fig.2}--\ref{Fig.4} are based on a statistical average over $10000$ trajectories of adaptive random compiler. At each step, the adaptive protocol uses expectation values with a statistical standard deviation of $0.1$. These settings are consistently applied to all three model Hamiltonians. In contrast, the results in Fig.~\ref{Fig.5} use a different approach, with expectation values evaluated exactly without statistical error.
	
	\subsection{Discrete-variable system}
	
	As a canonical discrete-variable system, the one-dimensional mixed-field Ising model (MFIM) extends the classical Ising framework through the simultaneous introduction of transverse and longitudinal magnetic fields. This extension enables a quantitative investigation of the emergence and evolution of quantum chaotic behavior, governed by the Hamiltonian:
	\begin{eqnarray}
		\hat{H}=-J\sum_{i=1}^{L}[\sigma_z^i\sigma_z^{i+1}+h_x\sigma_x^i+h_z\sigma_z^i],
	\end{eqnarray}
	where $\sigma_\alpha^i$,$\alpha\in x,y,z$ are the Pauli matrices acting on the $i$th site, $L$ is the length of the chain, $J$ is the Ising exchange of nearest neighbor spin $1/2$ and $h_{x/z}$ are the relative strengths of the transverse and longitudinal fields, respectively. The MFIM reduces to the transverse field Ising model (TFIM) at $h_z=0$, which can be exactly solved via the Jordan-Wigner transformation and describes free fermions. We consider periodic boundary conditions $\sigma_{L+1}=\sigma_1$.
	
	\begin{figure}[htbp]
		\includegraphics[width=\linewidth]{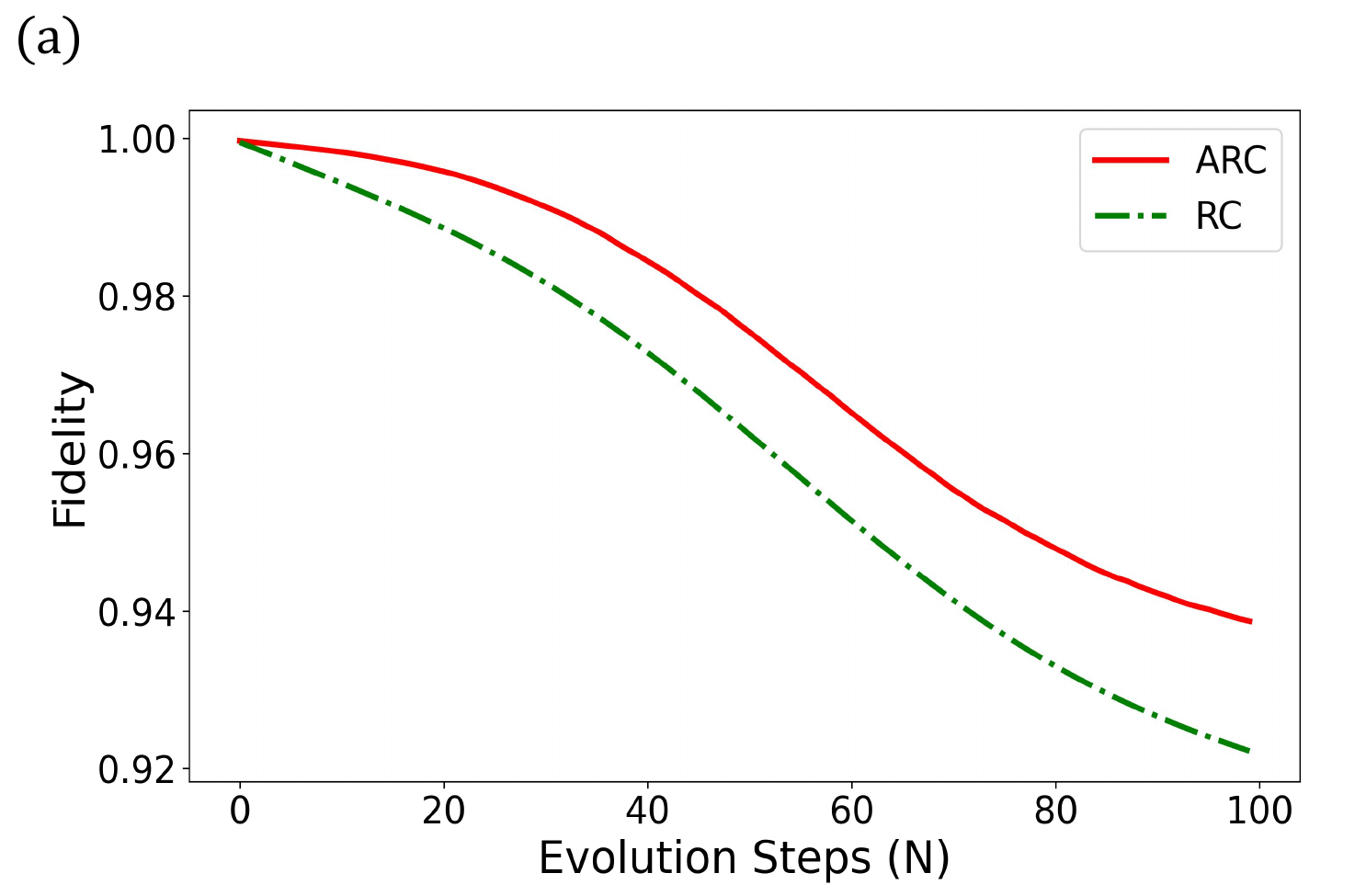}
		\includegraphics[width=\linewidth]{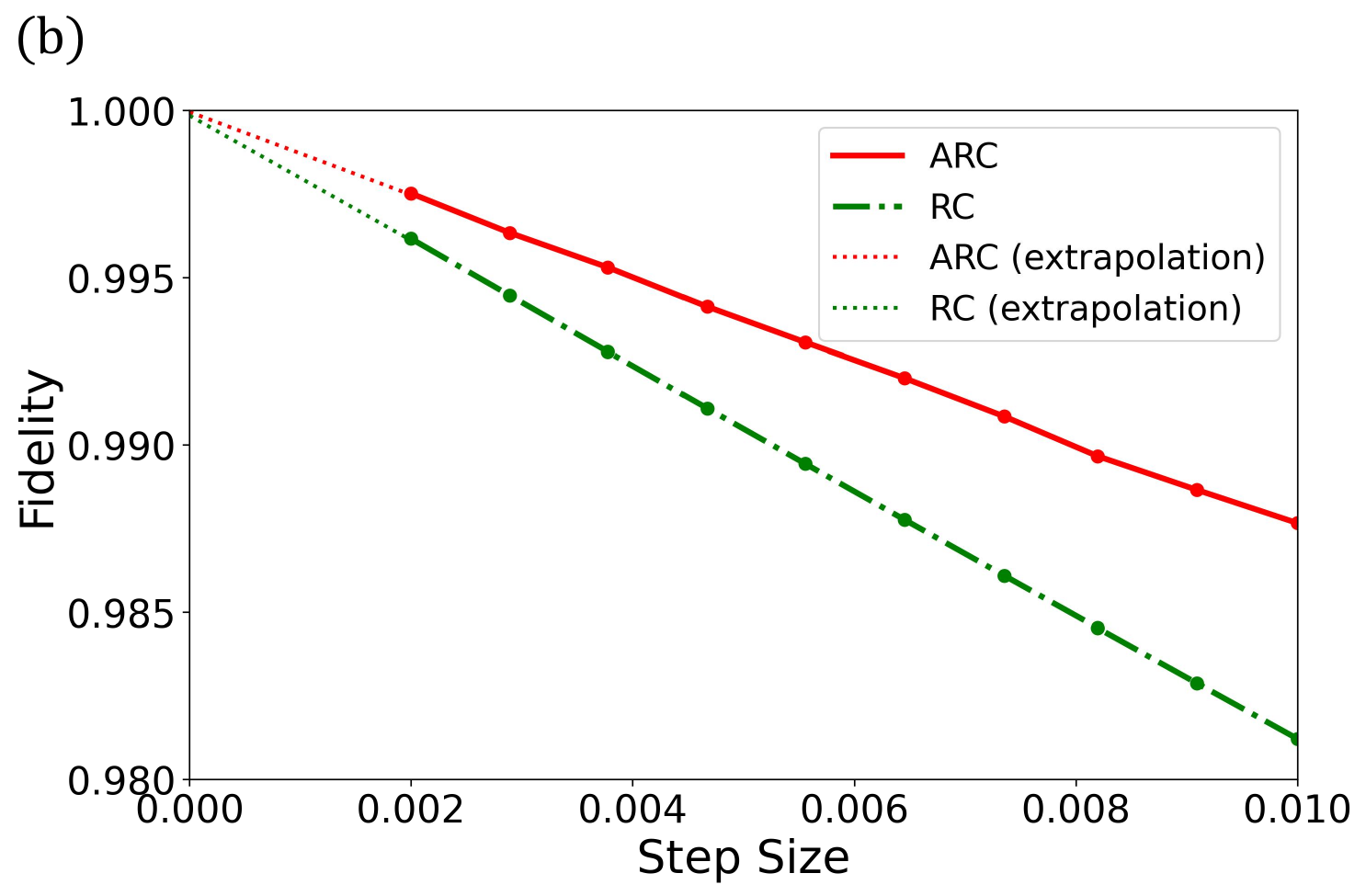}
		\caption{Fidelity comparison of the mixed-field Ising model Hamiltonian simulation. (a) Fidelity versus the number of evolution steps at a fixed step size $t/N=0.02$. (b) Fidelity as a function of step size with the total evolution time fixed at $t=1$. In both subfigures, the red solid line corresponds to the adaptive random compiler, while the green dot-dashed line shows results for the original random compiler protocol. The dotted line in (b) represents the fidelity extrapolated to zero step size. Other parameters are $L=4$, $J=1$, $h_x=0.5$, $h_z=0.3$, with initial state $\ket{0011}$.\label{Fig.2}}
	\end{figure}
	
	We test both random compiler and adaptive random compiler on the Hamiltonian, which is decomposed into three terms: $H_{zz}=-J\sum_{i=1}^{L}\sigma_z^i\sigma_z^{i+1}$, $H_{x}=-Jh_x\sum_{i=1}^{L}\sigma_x^i$ and $H_{z}=-Jh_z\sum_{i=1}^{L}\sigma_z^i$. This decomposition offers practical advantages, For the $H_x$ and $H_z$ terms, the unitary evolution operators within each group commute with one another, allowing all corresponding quantum gates to be applied in parallel. As for the $H_{zz}$ term, since it involves interactions between neighboring qubits, we can divide the terms into even and odd groups, where the operators within each group commute, thus enabling parallel gate execution as well. This parallelism is particularly important for near-term quantum devices with limited coherence times, as it helps mitigate the effects of noise and decoherence during simulation.
	
	Fig.~\ref{Fig.2} presents a comparison of fidelity between the states generated by each method and the exact time-evolved state in simulating the mixed-field Ising model Hamiltonian. In subfigure (a), we fix the step size at $t/N = 0.02$ and examine how the fidelity varies with the number of evolution steps. In subfigure (b), we fix the total evolution time at $t = 1$ and study the dependence of fidelity on the step size. In both cases, the red solid line denotes the performance of the adaptive random compiler, while the green dot-dashed line corresponds to the original random compiler protocol. Additionally, the dotted line in subfigure (b) depicts the extrapolated fidelity in the zero step size limit, demonstrating that both approaches asymptotically reach a fidelity of $1$, thereby validating their correctness. All other parameters are set to $L=4$, $J=1$, $h_x=0.5$, $h_z=0.3$, and an initial state $\ket{0011}$, representing half spins up and half down. As shown, although the fidelity decreases as the number of evolution steps and the step size increase in both methods, our strategy consistently yields better performance.
	
	\subsection{Continuous-variable system}
	
	Continuous-variable systems intrinsically possess infinite-dimensional Hilbert spaces, endowing them with natural advantages for certain simulation tasks~\cite{Marshall2015quantum,Abel2024simulating,Abel2025realtime}. However, the original random compiler protocol defines the Hamiltonian strength via the Schattern-$\infty$ norm, which causes difficulties when extending to continuous variable systems, while our method does not. Here we provide an example of a continuous-variable system to demonstrate this point.
	
	The driven Kerr oscillator exemplifies continuous-variable systems, characterizing a nonlinear optical cavity where a single-mode electromagnetic field interacts with a Kerr medium under external driving. Widely implemented in both quantum optical systems and superconducting circuits, its dynamics under rotating wave approximation are captured by the Hamiltonian:
	\begin{eqnarray}
		\hat{H}=\Delta \hat{a}^\dagger\hat{a}+\frac{K}{2}\hat{a}^\dagger\hat{a}^\dagger\hat{a}\hat{a}+\epsilon \left(\hat{a}+\hat{a}^\dagger\right),
	\end{eqnarray}
	where $\hat{a}$ and $\hat{a}^\dagger$ denote the annihilation and creation operators of the oscillator mode, respectively. The parameter $\Delta=\omega_0-\omega_d$ represents the detuning between the intrinsic frequency $\omega_0$ of the system and the driving field frequency $\omega_d$. The Kerr coefficient $K$ quantifies the strength of the nonlinear optical response, while $\epsilon$ characterizes the driving strength of the external classical field.
	
	\begin{figure}[htbp]
		\includegraphics[width=\linewidth]{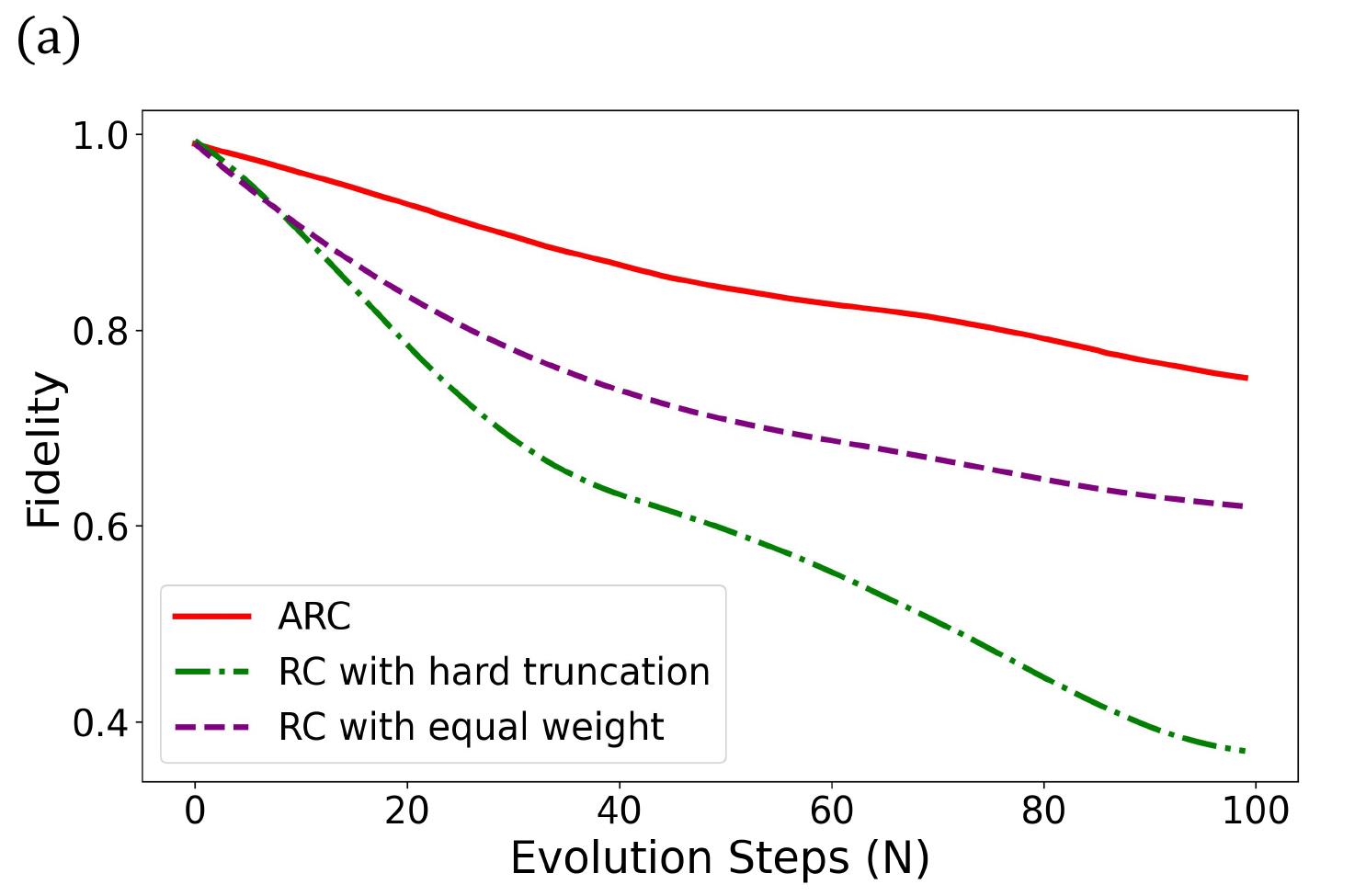}
		\includegraphics[width=\linewidth]{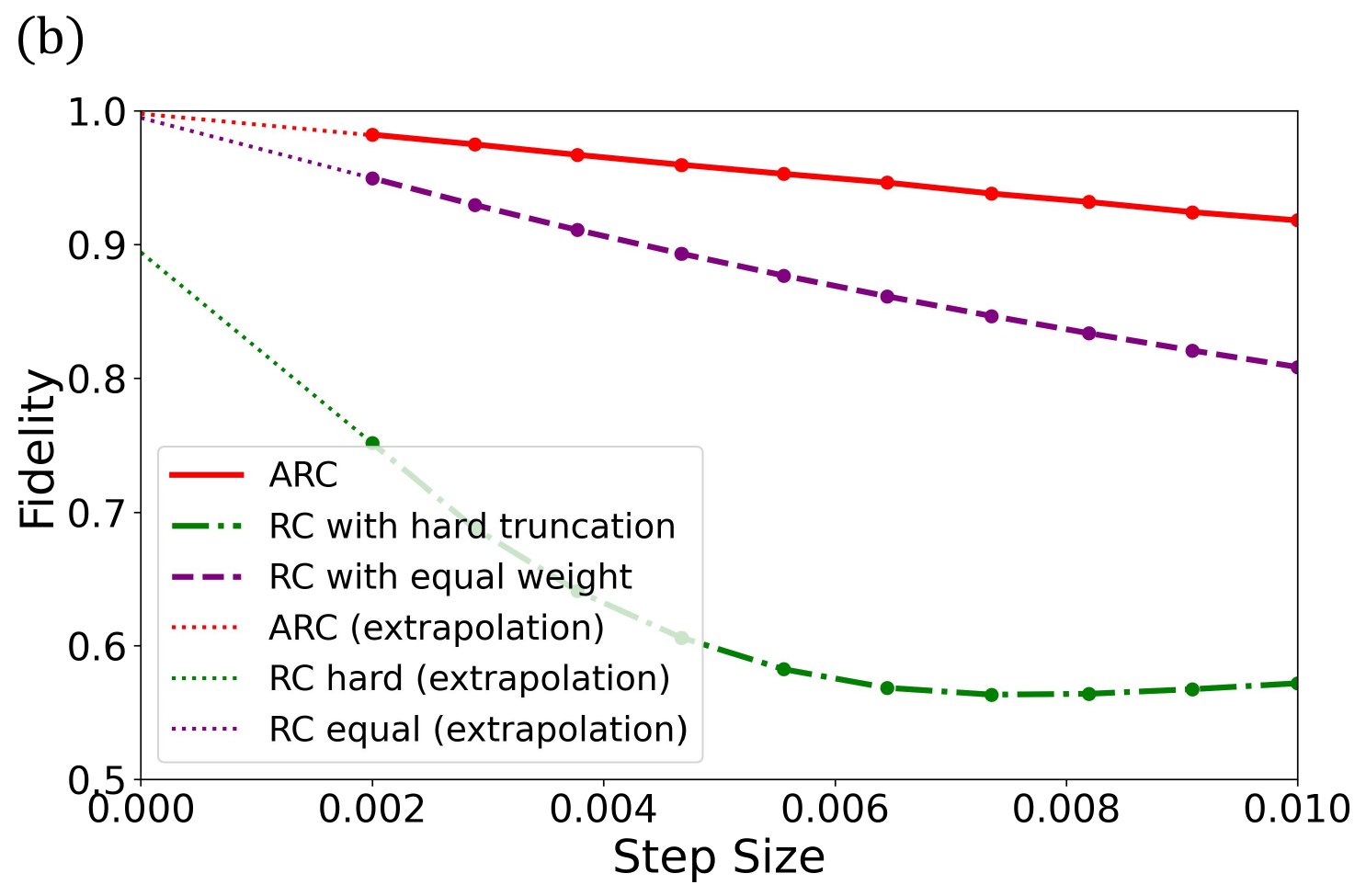}
		\caption{Driven Kerr oscillator Hamiltonian fidelity results. (a) Fidelity versus the number of evolution steps, with fixed step size $t/N=0.02$. (b) Fidelity dependence on step size for a fixed total evolution time $t=1$. The red solid line denotes our method, the green dot-dashed line indicates the original protocol with hard truncation, and the purple dashed line corresponds to the random compiler assigning equal weights to all Hamiltonian terms. The dotted line in (b) shows extrapolated fidelity at zero step size. The parameters are set as $\Delta=0.3$, $K=1$, $\epsilon=0.5$, initial state $(\ket{1}+\ket{5})/\sqrt{2}$, and Fock space truncation dimension $D=50$.\label{Fig.3}}
	\end{figure}
	
	We apply three methods to the driven Kerr oscillator model, whose Hamiltonian is partitioned into three terms: $\Delta\hat{a}^\dagger\hat{a}$, $K\hat{a}^\dagger\hat{a}^\dagger\hat{a}\hat{a}/2$ and $\epsilon( \hat{a}+\hat{a}^\dagger )$. These methods considered include the adaptive random compiler, the original randomized compiling protocol with hard truncation, and equal weight randomized compiling. In the case of hard truncation, the continuous-variable operators become bounded, which allows for the evaluation of operator strengths and the corresponding sampling probabilities. On the other hand, the equal weight scheme is adopted because all Hamiltonian terms are unbounded prior to truncation, making it infeasible to define relative strengths. Therefore, each term is sampled with equal probability to determine the applied quantum gate. 
	
	Fig.~\ref{Fig.3} shows fidelity performance in simulating the driven Kerr oscillator Hamiltonian. In subfigure (a), fidelity is plotted against the number of evolution steps with a fixed step size of $t/N = 0.02$. In subfigure (b), fidelity is shown as a function of the step size while keeping the total evolution time fixed at $t = 1$. In both subfigures, the red solid line denotes our method, the green dot-dash line represents the original protocol with hard truncation, and the purple dashed line corresponds to random compiler with equal weight for all terms. In addition, the dotted line in subfigure (b) represents the extrapolated fidelity as the step size approaches zero, showing that both methods yield fidelities approaching $1$, while the randomized compiling with hard truncation does not reach $1$ due to insufficient linearity in the curve, which makes the extrapolation less accurate. All other parameters are set to $\Delta=0.3$, $K=1$, $\epsilon=0.5$, an initial state $(\ket{1}+\ket{5})/\sqrt{2}$, and a Fock space truncation dimension $D=50$. Our approach again shows clear improvement over the baselines.
	
	\subsection{Hybrid-variable systems}
	
	Hybrid variable systems combine the controllability of discrete variables with the expressive power of continuous variables, enabling a broader range of applications~\cite{Zhang2020protocol,Zhang2021continuous,Andersen2015hybrid,Sabatini2024hybrid,Lepp2025quantum}. However, when applying randomized compiling techniques to such systems, the issue of defining the Hamiltonian strength becomes more severe, as hybrid-variable systems involve both bounded and unbounded operators. Therefore, we also present a simulation result for a hybrid variable system to illustrate this point.
	
	\begin{figure}[htbp]
		\includegraphics[width=\linewidth]{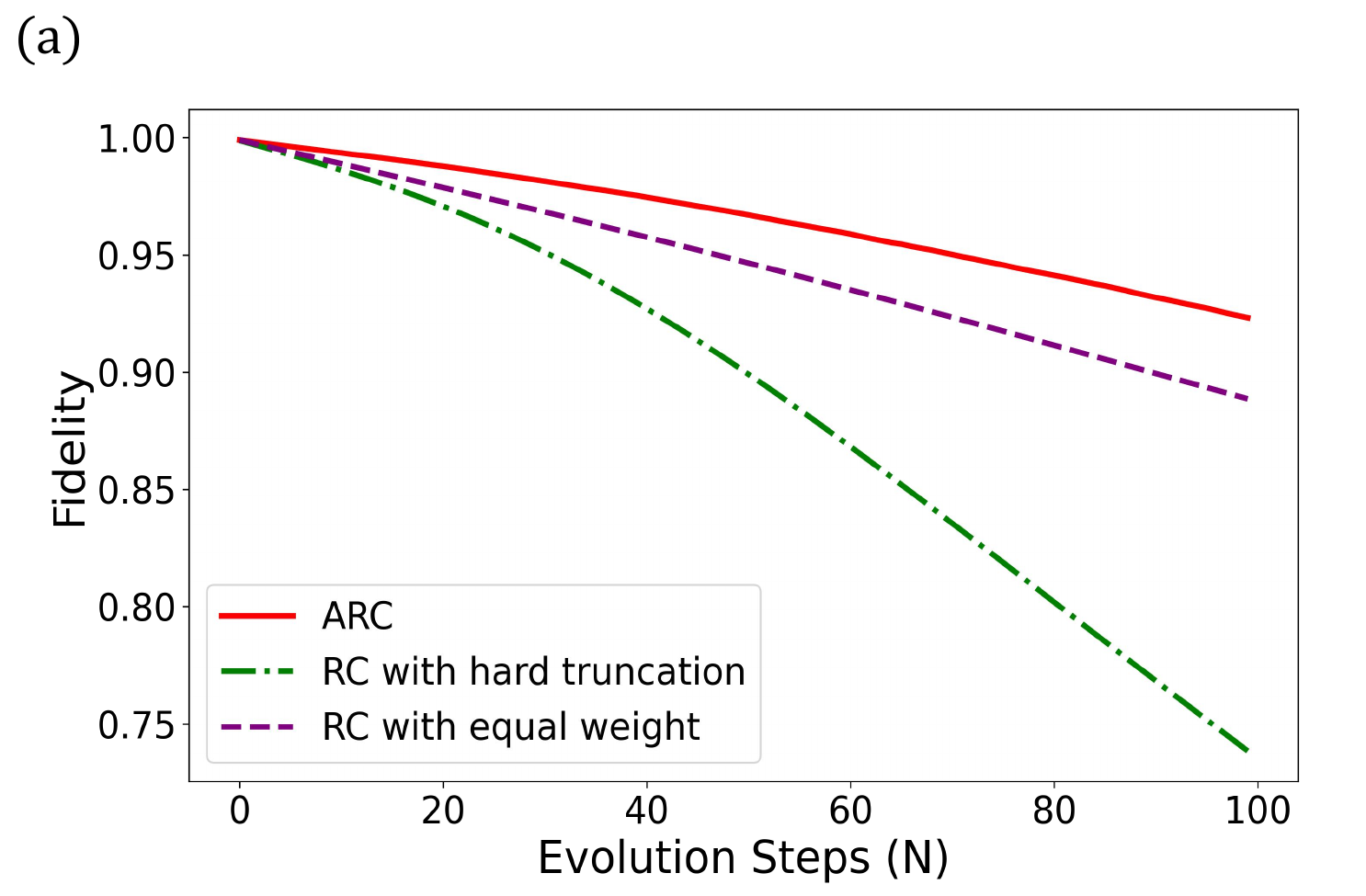}
		\includegraphics[width=\linewidth]{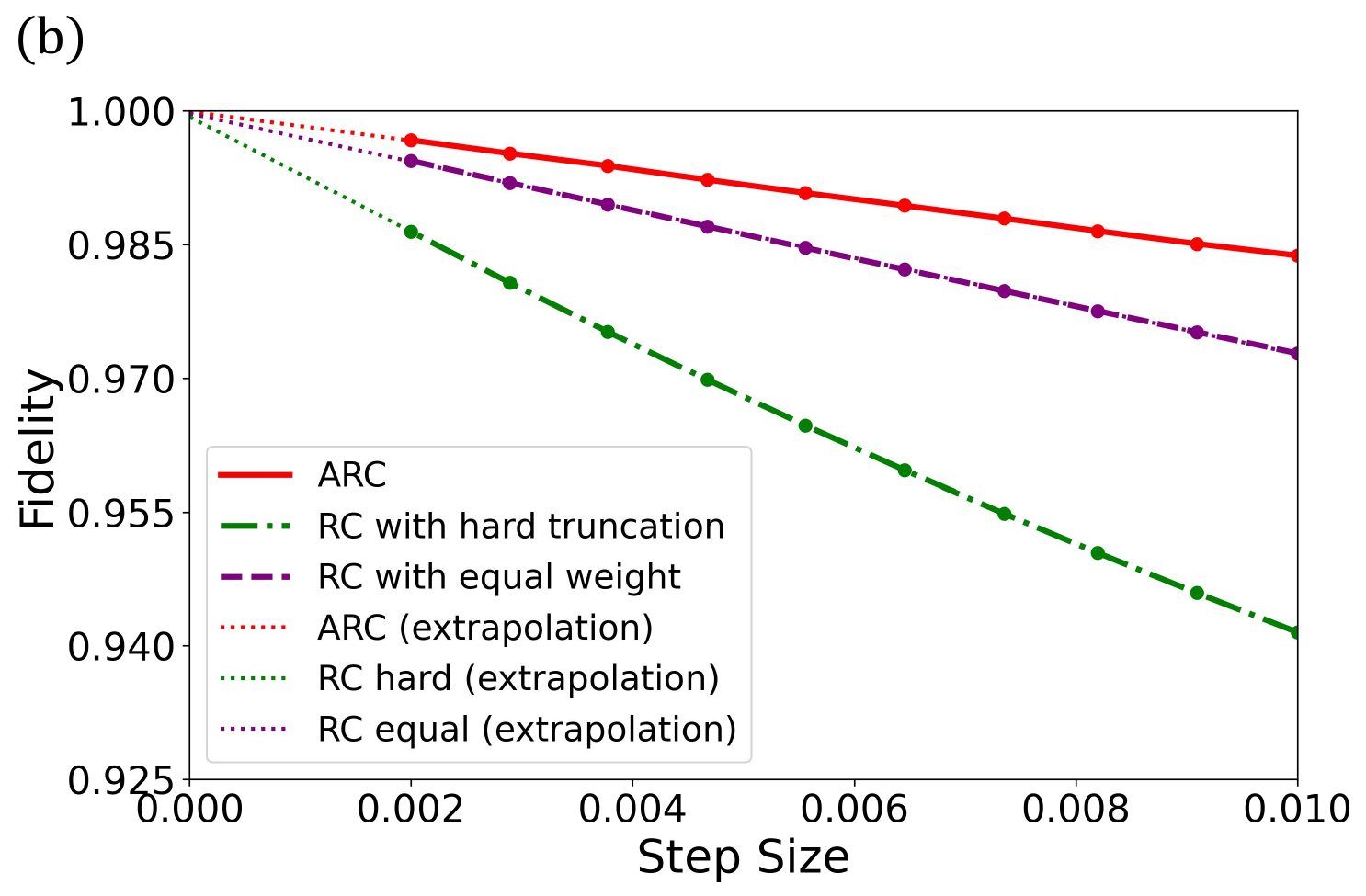}
		\caption{Quantum Rabi model Hamiltonian fidelity comparison. (a) Fidelity as a function of the number of evolution steps with fixed step size $t/N=0.02$. (b) Fidelity versus step size at total evolution time $t=1$. In both subfigures, the red solid line shows our method results, the green dot-dashed line corresponds to the original hard truncation protocol, and the purple dashed line depicts the random compiler with equal weighting. The dotted line in (b) represents the zero step size extrapolation. Parameters used are $\omega=1$, $\Omega=1$, $g=0.2$, initial state $(\ket{2,0}+\ket{5,0})/\sqrt{2}$, and truncation dimension $D=50$.\label{Fig.4}}
	\end{figure}
	
	Representing the fundamental hybrid-variable system in quantum physics, the Rabi model provides the minimal framework for studying light-matter interaction. It couples a discrete two-level system with a continuous bosonic field mode, with the complete interaction described by:
	\begin{eqnarray}
		\hat{H}=\omega\hat{a}^\dagger\hat{a}+\frac{\Omega}{2}\sigma_z+g\left(\hat{a}+\hat{a}^\dagger\right)\sigma_x
	\end{eqnarray}
	where $\hat{a}$ and $\hat{a}^\dagger$ denote the photon annihilation and creation operators, respectively. $\sigma_x$ and $\sigma_z$ are the Pauli matrices. The parameter $\omega$ represents the field frequency, $\Omega$ is the transition frequency of the two-level system, and $g$ denotes the light-matter coupling strength.
	
	We examined the quantum Rabi model, whose Hamiltonian was split into three parts: $\omega\hat{a}^\dagger\hat{a}$, $\Omega\sigma_z/2$ and $g(\hat{a}+\hat{a}^\dagger)\sigma_x$. 
	
	\begin{figure}[htbp]
		\includegraphics[width=\linewidth]{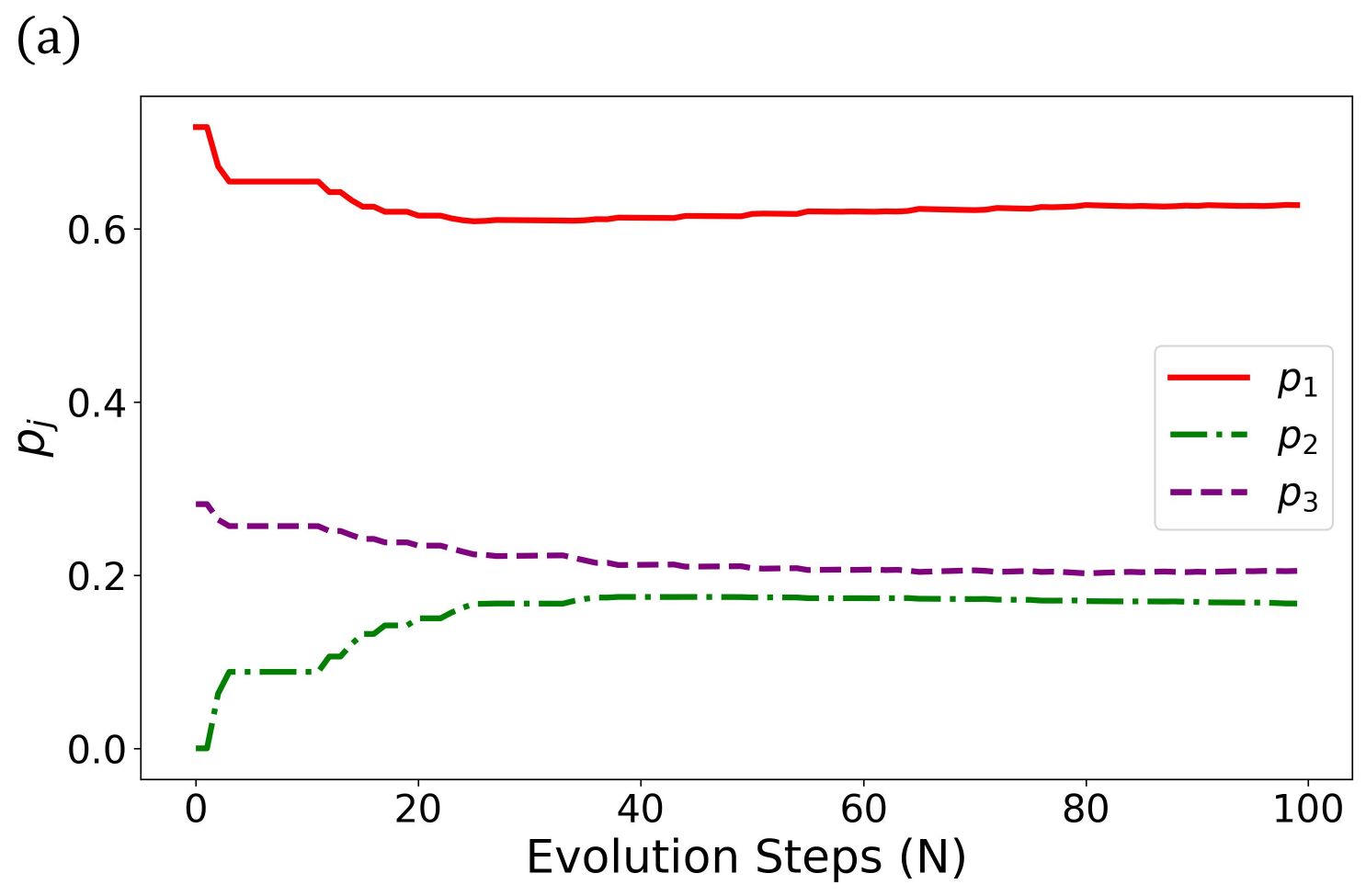}
		\includegraphics[width=\linewidth]{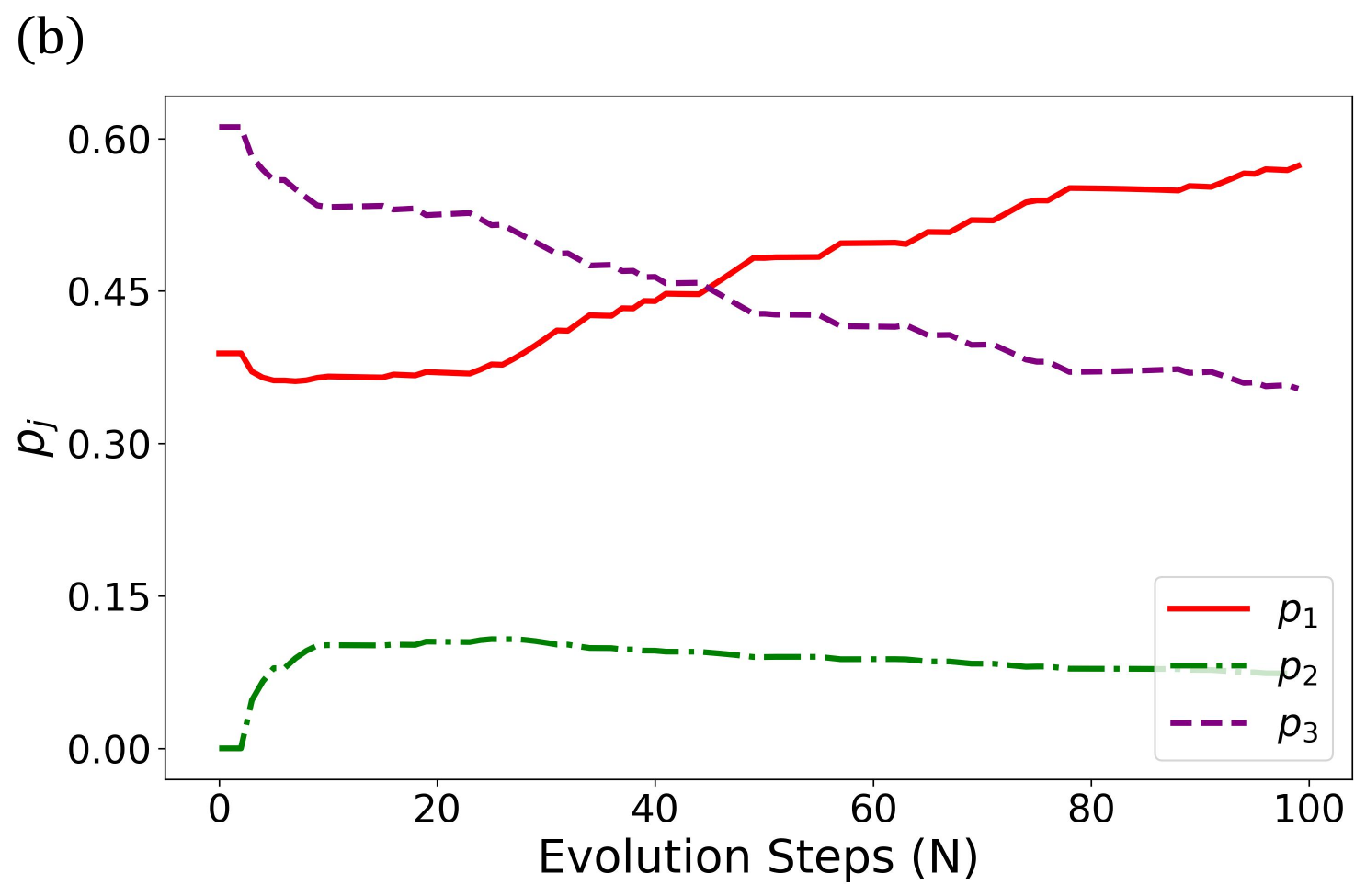}
		\caption{Dynamic adjustment of the probability distribution during the simulation of the Rabi model Hamiltonian using the adaptive random compiler.  Subfigures (a) and (b) correspond to coupling strengths $g=0.2$ and $g=0.8$, respectively, while other parameters are fixed as $D=50$ (the Fock space truncation dimension), $\omega=1$, $\Omega=1$ and $t/N=0.02$. The initial state in both cases is $(\ket{2,0}+\ket{5,0})/\sqrt{2}$. The red solid line, green dot-dash line and purple dashed line represent the sampling probabilities $p_1$, $p_2$ and $p_3$ corresponding to the three Hamiltonian terms, respectively.\label{Fig.5}}
	\end{figure}
	
	Fig.~\ref{Fig.4} shows fidelity performance in simulating the quantum Rabi model Hamiltonian. In subfigure (a), fidelity is plotted against the number of evolution steps with a fixed step size of $t/N = 0.02$. In subfigure (b), fidelity is shown as a function of the step size while keeping the total evolution time fixed at $t = 1$. In both subfigures, the red solid line denotes our method, the green dot-dash line represents the original protocol with hard truncation, and the purple dashed line corresponds to random compiler with equal weighting for all terms. In addition, the dotted line in subfigure (b) depicts the extrapolated fidelity in the zero step size limit, demonstrating that all three approaches asymptotically reach a fidelity of 1. All other parameters are set to $\omega=1$, $\Omega=1$, $g=0.2$, an initial state $(\ket{2,0}+\ket{5,0})/\sqrt{2}$, and a Fock space truncation dimension $D=50$. As illustrated in Fig.~\ref{Fig.4}, the adaptive compiler again outperforms both the original protocol with hard truncation and the equal-weighted version.
	
	To more intuitively illustrate the dynamic weight-tuning process, we record the optimal probability distribution at each step and plot it in Fig.~\ref{Fig.5}. The coupling strength $g$ is set to $0.2$ in subfigure (a) and $0.8$ in subfigure (b), while all other configurations remain unchanged from the previous settings. The red solid line, green dot-dash line and purple dashed line represent the sampling probabilities $p_1$, $p_2$ and $p_3$ corresponding to the three Hamiltonian terms, respectively.
	
	From Fig.~\ref{Fig.5}, we can observe two different modes of probability distribution modification. In subfigure (a), the distribution is updated at each step, which is reflected in the fluctuations of the curve. It should be noted that the severity of the fluctuation is also affected by the measurement accuracy of $\left|\left|D_{jj}(\rho)\right|\right|$. To eliminate visual disturbance caused by finite measurement accuracy, the expectation values used in the adaptive protocol at each step are evaluated exactly, without statistical error. In contrast, subfigure (b) not only exhibits fluctuations but also shows that the dominant Hamiltonian term changes as the evolution proceeds.
	
	\section{conclusion}\label{section 4}
	In summary, we have proposed an adaptive random compilation strategy for Hamiltonian simulation that improves the accuracy by dynamically tuning the sampling weights of Hamiltonian terms based on real-time feedback from low-order moment measurements. In parallel, our approach has addressed the challenge of characterizing the strength of unbounded operators in continuous-variable systems, thereby extending the reach of random compiler to a broader class of quantum platforms. We have validated the method through three numerical simulations, covering discrete-variable, continuous-variable and hybrid-variable quantum systems. This work has improved and generalized the random compiler algorithm, enabling broader and more flexible applications in quantum simulation.
	
	\begin{acknowledgments}
		This work was supported by the National Natural Science Foundation of China (Grant No.12375013) and the Guangdong Basic and Applied Basic Research Fund (Grant No.2023A1515011460).
	\end{acknowledgments}
	
	\section*{DATA AVAILABILITY}
	The data that support the findings of this article are openly available~\cite{Data}; embargo periods may apply.
	
	\appendix
	\setcounter{equation}{0}
	\renewcommand{\theequation}{\Alph{section}\arabic{equation}}
	
	\section{state-dependent circuit depth}
	Let $\Phi$ denote the exact single-step evolution channel, and $\tilde{\Phi}$ its approximate counterpart. For an initial state $\rho_0$, the states obtained after the exact and approximate evolutions for $N$ steps are $\rho_N = \Phi^N(\rho_0)$ and $\tilde{\rho}_N = \tilde{\Phi}^N(\rho_0)$, respectively. By applying the standard telescoping expansion, the difference between these two states can be expressed as
	\begin{eqnarray}
		&&\big\|(\tilde{\Phi}^N-\Phi^N)(\rho_0)\big\|\nonumber\\
		&&=\left\|\sum_{i=0}^{N-1}\tilde{\Phi}^{N-1-i}\!\left[(\tilde{\Phi}-\Phi)\Phi^i(\rho_0)\right]\right\|.
	\end{eqnarray}
	where $||A||=\sqrt{\mathrm{Tr} (A^\dagger A)}$ denotes the Hilbert--Schmidt norm. Using the triangle inequality to the summation yields
	\begin{eqnarray}
		&&\big\|(\tilde{\Phi}^N-\Phi^N)(\rho_0)\big\|\nonumber\\
		&&\le \sum_{i=0}^{N-1}\Big\|\tilde{\Phi}^{N-1-i}\!\left[(\tilde{\Phi}-\Phi)\Phi^i(\rho_0)\right]\Big\|.
	\end{eqnarray}
	When $\tilde{\Phi}$ is unital, it is contractive with respect to the Hilbert--Schmidt norm~\cite{Perez2006contractivity}. Consequently, the preceding factors $\tilde{\Phi}^{N-1-i}$ in the telescoping expansion do not enlarge the norm, allowing us to bound the total deviation as
	\begin{eqnarray}
		\big\|(\tilde{\Phi}^N-\Phi^N)(\rho_0)\big\|&\leq&\sum_{i=0}^{N-1}\big\|(\tilde{\Phi}-\Phi)(\Phi^i(\rho_0))\big\|\nonumber\\
		&=&\sum_{i=0}^{N-1}\big\|\tilde{\Phi}(\rho_i)-\Phi(\rho_i)\big\|.
		\label{Eq.22}
	\end{eqnarray}
	where $\rho_i:=\Phi^i(\rho_0)$ denotes the exact state at the $i$-th step of the evolution.

	This framework offers a unified approach to analyzing the state-dependent error accumulation for both the Trotter and random compiler methods. We then substitute the respective single-step approximations to examine the resulting errors.

	For the first-order Trotter approach~\cite{Trotter1959product}, the deviation from the exact evolution up to second order, acting on an arbitrary state $\sigma$, is
	\begin{eqnarray}
		\big\|(\tilde{\Phi}-\Phi)(\sigma)\big\|\approx\frac{ t^2}{2N^2}\left\|\left(\sum_{j<k}[\mathcal{L}_j,\mathcal{L}_k]\right)(\sigma)\right\|.\nonumber\\
	\end{eqnarray}
	Substituting into Eq.~\eqref{Eq.22} gives 
	\begin{eqnarray}
		\big\|(\tilde{\Phi}^N-\Phi^N)(\rho_0)\big\|&\leq& \frac{t^2}{2N^2} \sum_{i=0}^{N-1} \left|\left| \sum_{j<k} [\mathcal{L}_j, \mathcal{L}_k](\rho_i) \right|\right| \nonumber\\
		&=& \frac{t^2}{2 N^2} \sum_{i=0}^{N-1} \left|\left| \sum_{j<k} [\mathcal{L}_j, \mathcal{L}_k](\rho_i) \right|\right| \nonumber\\
		&=& \frac{t^2}{2 N} \, \overline{\left|\left| \sum_{j<k} [\mathcal{L}_j, \mathcal{L}_k](\rho_i) \right|\right|}.
	\end{eqnarray}
	where $\overline{(\cdot)}$ denotes the average over the $N$ steps. Accordingly, the total error depends on the average contribution of the commutators over the intermediate states. Here, $N$ refers to the number of Trotter steps. As each step involves $L$ gate layers, the circuit depth must counted as $NL$.

	For the random compiler (RC) and the adaptive random compiler (ARC), the second-order deviation from the exact evolution follows the same analytical derivation. The only difference lies in the choice of the probability distribution $p_j$. Therefore, we keep $p_j$ general in the derivation. For an arbitrary state $\sigma$, the deviation can be expressed as
	\begin{eqnarray}
		\big\|(\tilde{\Phi}-\Phi)(\sigma)\big\| \approx \frac{t^2}{2N^2}\left\|\left(\mathcal{L}^2(\sigma) - \sum_j \frac{\mathcal{L}_j^2(\sigma)}{p_j}\right)\right\|.\nonumber\\
	\end{eqnarray}
	By substituting into Eq.~\eqref{Eq.22}, we obtain
	\begin{eqnarray}
		&&\big\| (\tilde{\Phi}^N - \Phi^N)(\rho_0) \big\|\nonumber\\
		&&\le\frac{t^2}{2N^2}\sum_{i=0}^{N-1}\left|\left|\mathcal{L}^2(\rho_i)-\sum_{j}\frac{\mathcal{L}_j^{2}(\rho_i)}{p_j}\right|\right|\nonumber\\
		&&\leq\frac{t^2}{2N^2}\sum_{i=0}^{N-1}\left(\left|\left|\mathcal{L}^2(\rho_i)\right|\right|+\sum_{j}\frac{\left|\left|\mathcal{L}_j^{2}(\rho_i)\right|\right|}{p_j}\right)\nonumber\\
		&&=\frac{t^2}{2N}\overline{\left(\left|\left|\mathcal{L}^2(\rho_i)\right|\right|+\sum_{j}\frac{\left|\left|\mathcal{L}_j^{2}(\rho_i)\right|\right|}{p_j}\right)}.
	\end{eqnarray}
	where $\overline{(\cdot)}$ denotes the average over the $N$ steps. The total error depends on the average contributions of the collective generator $\mathcal{L}^2$ and the weighted individual generators $\mathcal{L}_j^2/p_j$ over the intermediate states.

	Finally, by substituting $p_j = \frac{||H_j||_{\infty}}{\sum_{k}||H_k||_{\infty}}=\frac{||H_j||_{\infty}}{\lambda}$ for RC and $p_j = \frac{\sqrt{||\mathcal{L}_j^2(\rho_i)||}}{\sum_{k}\sqrt{||\mathcal{L}_k^2(\rho_i)||}}$ for ARC, we obtain the corresponding final results, where $\|\cdot\|_{\infty}$ denotes the Schatten-$\infty$ norm, which is equal to the largest singular value of an operator.
	
	For the random compiler~\cite{Campbell2019random},
	\begin{eqnarray}
		&&\big\| (\tilde{\Phi}^N - \Phi^N)(\rho_0) \big\|\nonumber\\
		&&\leq
		\frac{t^2}{2N}\overline{
			\left(
			\left|\left|\mathcal{L}^2(\rho_i)\right|\right|
			+\lambda\sum_{j}
			\frac{
				\left|\left|\mathcal{L}_j^{2}(\rho_i)\right|\right|
			}{||H_j||_{\infty}}
			\right)
		}.
	\end{eqnarray}
	
	For the adaptive random compiler proposed in this work,
	\begin{eqnarray}
		&&\big\| (\tilde{\Phi}^N - \Phi^N)(\rho_0) \big\|\nonumber\\
		&&\leq
		\frac{t^2}{2N}\overline{
			\left(
			\left|\left|\mathcal{L}^2(\rho_i)\right|\right|
			+\left(\sum_{j}
			\sqrt{||\mathcal{L}_j^2(\rho_i)||}\right)^2
			\right)
		}.\nonumber\\
	\end{eqnarray}

	\bibliography{ADARC}
	
\end{document}